          [2004/04/02 1.0 (PWD)]
\documentclass[a4paper,twocolumn]{esapub} % European paper
\usepackage{natbib}
\usepackage{graphicx}
\title{Search for $^{44}$Ti gamma-ray line emission from GRO J0852-4642 with INTEGRAL/SPI}
\author[1]{A. von Kienlin}
\affil[1]{Max-Planck-Institut f\"ur extraterrestrische Physik, PO Box 1312, 85741 Garching, Germany}
\author[2]{D. Atti\'e}
\author[2]{S. Schanne}
\author[2]{B. Cordier}
\affil[2]{CEA/Saclay, DAPNIA/Service d'Astrophysique, F-91191 Gif sur Yvette, France}
\author[1]{R. Diehl}
\author[1]{A. F. Iyudin}
\author[1]{G. G. Lichti}
\author[3]{J.-P. Roques}
\author[1]{V. Sch\"onfelder}
\author[1]{A. Strong}
\affil[3]{Centre d'Etudes Spatiales des Rayonnements, 9, avenue du Colonel Roche, 31028 Toulouse, France}

\begin{document}

\keywords{gamma rays: observations, nucleosynthesis, Space telescope: SPI/INTEGRAL}

\maketitle

\begin{abstract}

The gamma-ray source GRO J0852-4642, discovered by COMPTEL, is a possible counterpart of the supernova remnant RX %%@
J0852-4622. 
Detection of radioactive decay from $^{44}$Ti nuclei  would proof it to be the youngest and nearest supernova remnant %%@
known so far.
During the first year of {\mbox {\it INTEGRAL}} core program, the Vela region was observed twice 
in all for more than 2000~ks. Among other nucleosynthesis studies, one of the most
important scientific goals of this observation is the detection of $^{44}$Ti gamma-ray lines
expected at 68 keV, 78 keV and 1157 keV. For this purpose the {\mbox {\it INTEGRAL}}
Spectrometer (SPI), with its very high energy resolution thanks to its germanium detector
camera is the key instrument, permitting a precise determination of gamma-ray line
intensities and profiles. 
The upper limit for the 78.4 keV $^{44}$Ti gamma-ray line emission derived from the first analysis 
is $1.1 \times 10^{-4}\, \rm \gamma \, cm^{-2}\, s^{-1}$. This value is mainly dominated by systematic 
uncertainties in the treatment of the instrumental background. By accumulating more observation time 
in the next years of the mission and by improving the background understanding, a reliable $^{44}$Ti flux 
for GRO J0852-4642 or an upper limit which constrains the COMPTEL flux can be expected. 
\end{abstract}

\section{Introduction}

One of the main scientific topics of the {\mbox {\it INTEGRAL}} mission is the investigation of
stellar nucleosynthesis (Hydrostatic burning, supernovae, novae) \citep{2003A&A...411L...1W}.
In this context the spectrometer SPI is important, due to its capability to resolve nuclear 
lines with its high energy resolution \citep{2003A&A...411L..63V,2003SPIE.4851.1132S,2003A&A...411L..71A}. The %%@
measurement of  $\gamma$-ray line intensities, line profiles 
and line shifts is one of SPI's main goals \citep{2000AIPC..510...54S}.

Only few isotopes exist which are accessible to  $\gamma$-ray astronomy for probing cosmic nucleosynthesis %%@
\citep{1998PASP..110..637D}. Especially $^{44}$Ti is an important probe for the investigation of supernovae (SNe) and %%@
their young remnants (SNR). 
$^{44}$Ti decays via $^{44}$Sc to $^{44}$Ca:

 \begin{math}
 ^{44}\rm{Ti}\left[ \frac{67.9\,\rm{keV,} \; 78.4 \rm\,{keV}}{\tau \,= \, 86.0 \,\pm\, 0.7 \,\rm{yr}}\right] %%@
\rightarrow \; ^{44}\rm{Sc}\left[ \frac{1157.0\rm\,{keV}}{\tau \,= \,5.67 \,\rm{h} }\right]\rightarrow \; %%@
^{44}\rm{Ca}
 \end{math}
 
This decay chain is driven by the decay-time of $^{44}$Ti. The 67.9 keV and 78.4 keV transitions form a cascade and %%@
have the same probability (99\%) as the 1157 keV transition, thus a similar flux of all three lines is expected. The %%@
mean life-time   of $^{44}$Ti was for a long time very uncertain, but has been settled a few years ago to a current %%@
"best value" of $86.0 \pm 0.7$ ~yr (1 $\sigma$ error)  [half-life time: 59.6 $\pm$ 0.5 yr], from more accurate %%@
laboratory measurements \citep{1998PhRvC..57.2010N,1998PhRvL..80.2550A,1998PhRvL..80.2554G,1999PhRvC..59..528W}. 
$^{44}$Ti decays through electron capture, therefore its decay lifetime depends on the ionization state: in the %%@
extreme case of a fully-ionized nucleus the $^{44}$Ti isotope is stable. This effect, if present, might introduce %%@
some uncertainty in the estimation of the $^{44}$Ti-mass ejected by the SN %%@
\citep{1999A&A...346..831M,2001ApJ...563..828L}. 

The discovery of 1157 keV $^{44}$Ti line emission from the historic youngest-known Galactic SNR Cas A by COMPTEL %%@
\citep{1994A&A...284L...1I} confirmed that $^{44}$Ti is indeed produced in core collapse SNe explosions. $^{44}$Ti is %%@
expected to be produced in all types of SNe, although with very different yields, depending on the mass-cut between a %%@
few 10$^{-5}$ and 10$^{-4}$ M$_{\odot}$ for the most frequent SNe of type II and Ib %%@
\citep{1995ApJS..101..181W,1996ApJ...460..408T}. $^{44}$Ti provides an excellent diagnostic of the still not %%@
completely understood core-collapse SN-explosion mechanism itself. According to theoretical models, $^{44}$Ti is %%@
produced through explosive Si burning in an  $\alpha$-rich freeze-out and thus originating from the innermost ejected %%@
material. $^{44}$Ti thus probes deep into the interior of the exploded star (similar to radioactive $^{56}$Ni). 
Its yield is sensitive to the asymmetries of the collapse and explosion, possibly caused by stellar rotation %%@
\citep{1998ApJ...492L..45N}. 

The recently-discovered SNR RXJ0852-4622 \citep{1998Natur.396..141A}, with $\sim$\,2$^\circ$ diameter, an age %%@
probably much below that of the Vela SNR, and a large distance uncertainty, from 0.2 to 1 kpc, is the object of our %%@
investigation. {\mbox {\it INTEGRAL}} observations have the potential to resolve the distance uncertainty from a test %%@
for $^{44}$Ti emission and from a study of the 1.8 MeV gamma-ray line shape expected to be broadened and produced by %%@
the decay of fast moving $^{26}$Al nuclei possibly released by this SNR, compared to the line shape (narrow) produced %%@
by slow moving $^{26}$Al nuclei which form a diffuse background in the Vela region.

The flux of the $^{44}$Ti line is given by
 
\begin{equation}
\Phi = \frac{1}{4 \pi \cdot d^2} \cdot \frac{Y_{\rm Ti}}{m_{\rm Ti}\cdot \tau_{\rm Ti}}\cdot e^{-\frac{t}{\tau_{\rm %%@
Ti}}},
\label{ti44yield}
\end{equation}

with $d$ as the distance of the supernova, $Y_{\rm Ti}$ as the mass yield of $^{44}$Ti, $m_{\rm Ti}$ as the atomic %%@
mass of $^{44}$Ti,  $\tau_{\rm Ti}$ as the  $^{44}$Ti mean life-time   and $t$ as the time elapsed since the %%@
supernova explosion.
Obviously, for any observed flux there is some ambiguity in the age and distance parameters of the source. This is %%@
where other astronomical observations must join in. 
Ice-core dating of nearby supernovae had been proposed for such purpose already in the 1970s. It was claimed that the %%@
measurement of nitrate abundance in Antarctic ice cores shows evidence for a correlation with known recent nearby %%@
supernovae \citep{1979Natur.282..701R,2000APh....14....1B}, due to photochemical reactions in the atmosphere. 

From the observations of X-rays, the age and distance of the newly-discovered SNR RXJ0852-4622 has been a matter of %%@
debate. From its size and spectrum, values well below 1 kpc were thought plausible \citep{1998Natur.396..141A}, while %%@
comparisons of absorption column data with the Vela SNR require distances well beyond the Vela SNR (currently placed %%@
at 250 pc; \cite{1999ApJ...515L..25C,2001ApJ...548..814S}). From an identification of RXJ0852-4622 with the $^{44}$Ti %%@
source, a combination of arguments led to a most plausible parameter set of 200 pc for the distance, and 680 years %%@
for the age of the SN \citep{1999A&A...350..997A}. Other tracers \citep{2002Atmos...64..669I} also support a SN event %%@
around 1320, the occurance time of the COMPTEL GRO J0852-4622 event estimated from $^{44}$Ti measurements %%@
\citep{1998Natur.396..142I}.
If this small distance value applies, an interesting perspective for {\mbox {\it INTEGRAL}} is opened from its %%@
capability to detect $^{26}$Al radioactive isotopes expected to be released by this SNR. Indeed, core-collapse %%@
supernovae are believed to be common sources of $^{26}$Al \citep{1996PhR...267....1P}, and could be visible with %%@
{\mbox {\it INTEGRAL}} as individual sources out to distances of a few 100 pc, assuming standard model yields, due to %%@
the long $^{26}$Al lifetime of $\sim$~10$^6$ y. Indeed, COMPTEL's $^{26}$Al survey shows a peak of emission towards %%@
this direction, which could be largely due to this source 
\citep{1999A&A...350..997A}.
First results of a search for $^{26}$Al emission in the Vela region with {\mbox {\it INTEGRAL}}/SPI are shown %%@
elsewhere \citep{2004ESA...Maurin}. A measurement of the $^{26}$Al line width in the Vela region, expected to be %%@
performed by SPI, will be crucial for the interpretation of the nature of the SNR {\mbox RX J0852-4622}.

The detection of 1157 keV $^{44}$Ti emission \citep{1998Natur.396..142I} has shown that it is possible to discover a %%@
supernova remnant, which was undetected in other wavelengths before, by the detection of  $\gamma$-ray lines. %%@
Although the $^{44}$Ti detection is only marginal \citep{2000AIPC..510...54S}, it is for the first time that a  %%@
$\gamma$-ray line observation triggered the discovery of a new supernova remnant, through X-rays measured with ROSAT. 
A confirmed $^{44}$Ti  $\gamma$-ray line emission for GRO J0852-4642 will help also to study the Galactic supernovae %%@
rate. $^{44}$Ti is an excellent indicator for Galactic supernova explosions which occurred in the past few centuries %%@
(age  $\leq$ 1000 yr), even in otherwise obscured regions. The Galaxy is, compared to other wavelength bands, %%@
practically transparent in the  $\gamma$-ray band. Therefore it will be possible to complement historical %%@
observations of galactic SNe through supernovae remnants revealed by their $^{44}$Ti  $\gamma$-ray line emission.

\begin{figure}[botttom]
\centering
\includegraphics[width=0.8\linewidth]{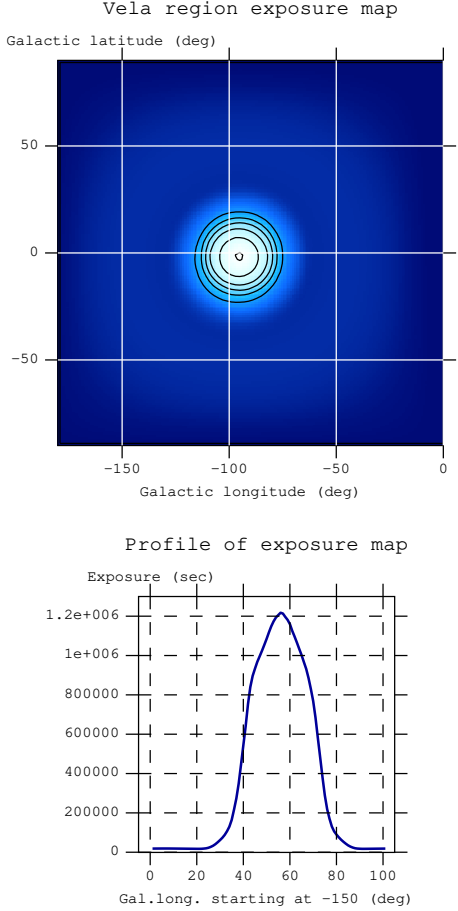}
\caption{Distribution map of the effective exposure of the Vela region. The location of GRO J0852-4642 coincides %%@
approximately with the maximum exposure of about 1200 ks.\label{fig:exposure}}
\end{figure}

\section{Observation of the Vela region}
\label{chapter:ObsVelaReg}

The Vela Region was observed twice during the {\mbox {\it INTEGRAL}} AO-I Core Program in 2003.

The first set of data with 1236 ks exposure time was recorded during the satellite orbits 81 to 88 
from June 12 until July 6. The 5$\times$5 dither pattern was centered in the Vela region at $\alpha_{\rm J2000} = %%@
8^{\rm h}\, 52^{\rm m} \,45.6^{\rm s}$ and $\delta_{\rm J2000} = -44^{\circ}\, 35'\, 07.2"$, $\sim 2^{\circ}$ off the %%@
position of the potential SNR GRO J0852-4642. 
During orbit 82 and 83 a large solar flare occurred with the consequence that the data of these orbits were left out %%@
for the analysis. 

The second set of data with 986 ks exposure time has been taken during the orbits 137 to 141 from November 27 until %%@
December 11. The 5$\times$5 dither pattern was centered on $\alpha_{\rm J2000} = 8^{\rm h}\, 27^{\rm m} \,54.7^{\rm %%@
s}$ and  $\delta_{\rm J2000} = -46^{\circ}\, 18'\, 18.4"$, $\sim 4^{\circ}$ off the position of GRO J0852-4642. In %%@
reaction to the Vela X-1 outburst on November 28 \citep{2003ATel..211....1K, 2004ESA...Staubert}, 
a 12.6 ks  observation with a hexagonal dither pattern centered on Vela X-1 ($\alpha_{\rm J2000} = 9^{\rm h}\, 2^{\rm %%@
m} \,6.9^{\rm s}$ and  $\delta_{\rm J2000} = -40^{\circ}\, 33'\, 7.2"$) was executed end of orbit 138 
on December 2.
Unfortunately one of SPI's germanium detectors (Ge-detector No. 2) failed at the beginning of orbit 140 on December %%@
6. 
The consequence is that the partly energy deposition in detectors which are involved in the same 
multiple detector event as the failed one will impose fake background events, especially in detectors 
which are located in the neighborhood of Ge-detector 2. A new detector response is required to account 
for the changed conditions. The current analysis, which is presented here, discards the data of orbit 140 and 141 
until the  new response is available. As it already happened during the first Vela observation, a strong solar flare %%@
occurred during the second period too, starting at the end of orbit 138 and lasting until the beginning of orbit 139. %%@
Due to the shorter duration of the activity, the recorded data were used for the analysis. 

Altogether the Vela Region was observed for 2235 ks. The data of the orbits 82, 83, 140, 141 and the satellite slews %%@
were discarded. Thus the on-target time used for the analysis  amounts 1540 ks. With the mean deadtime of all %%@
detectors and pointings of $\sim$ 11\% the effective exposure was 1370 ks. Fig.\,\ref{fig:exposure} shows the %%@
distribution map of the effective exposure.  With the selected data, only 1200 ks effective exposure (about 54\% of %%@
the observing time) are available for the spectral analysis of GRO J0852-4642.

\section{Sensitivity estimations}
\label{chapter:SensEst}

\begin{figure}
\centering
\includegraphics[width=1.0\linewidth]{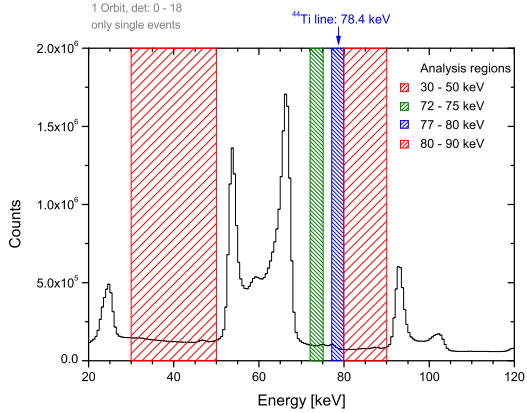}
\caption{Low energy part of typical spectrum of SPI's germanium camera. The spectrum comprises data from one orbit %%@
with the single events of all 19 detectors summed up after the energy calibration. \label{fig:spectrum}}
\end{figure}

The expectations to obtain a significant result for the 67.9 keV line is low, because SPI has at 67.9 keV a small %%@
sensitivity of ${\rm F}_{67.9} \approx 7 \times 10^{-5}\, \rm \gamma \, cm^{-2}\, s^{-1}$  caused by the strong %%@
background line complex between ~50 and ~70 keV (see Fig.\,\ref{fig:spectrum}). The narrow line sensitivity at the %%@
1157 keV line is ${\rm F}_{1157} \approx 2.3 \times 10^{-5}\, \rm \gamma \, cm^{-2}\, s^{-1}$, when adding the single %%@
and multiple detector events, it is about the same as the one at 78.4 keV of ${\rm F}_{78.4} \approx 2.5 \times %%@
10^{-5}\, \rm \gamma \, cm^{-2}\, s^{-1}$ (sensitivities taken from \cite{2003A&A...411L..91R}). However
at high energies SPI's sensitivity is deteriorated by strong Doppler line broadening which can be expected for this %%@
source. In the case of SPI the Doppler broadening of the 1157 keV line could exceed significantly the intrinsic %%@
energy resolution of the germanium detector, leading to a reduced sensitivity. For example a Doppler broadening of 35 %%@
keV, which corresponds to an expansion velocity of about 4600 km/s, only a significance of  $\sigma = 1.4$  can be %%@
obtained with an 3000 ks observation. 

\begin{figure}
\centering
\includegraphics[width=1.0\linewidth]{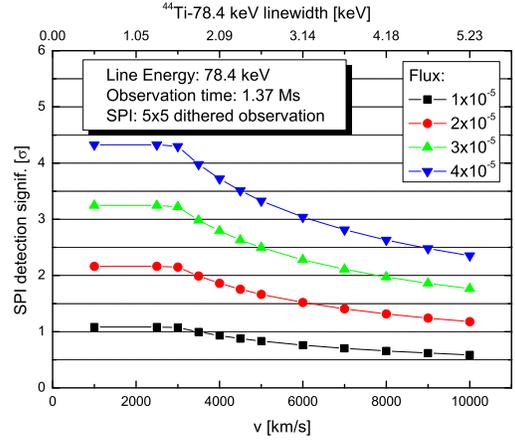}
\caption{Expected detection significances for the  $^{44}$Ti-line at 78.4 keV for source fluxes ranging from 1 to 4 $ %%@
\times 10^{-5}\, \rm \gamma \, cm^{-2}\, s^{-1}$. Observation conditions for SPI are 1370 ks observation time and a %%@
$5 \times 5$ dither mode.\label{fig:44Ti-sens}}
\end{figure}

Fig.\,\ref{fig:44Ti-sens}  summarize the expected significances for the detection of the 78.4 keV $^{44}$Ti-line with %%@
SPI for an 1370 ks  observation with a $5 \times 5 $ dither pattern ($\sim $16 days, which corresponds to the %%@
effective exposure as determined in Section\,\ref{chapter:ObsVelaReg}). The significances were derived by using the %%@
Observation Time Estimator (OTE) of ISOC for $^{44}$Ti-line fluxes in the range $F_{78.4 \rm \, keV} = (1 - 4) \times %%@
10^{-5} \rm \, \gamma \, cm^{-2} \, s^{-1}$. This range corresponds to the $^{44}$Ti excess measured with {\mbox %%@
COMPTEL} of $F_{1157 \rm \, keV} = (2 - 4) \times 10^{-5} \rm \, \gamma \, cm^{-2} \, s^{-1}$ from GRO J0852-4642 %%@
\citep{2000AIPC..510...54S}.
The expected significances are plotted for expansion velocities of the SNR between 1,000 and $10,000 \rm \, km\, %%@
s^{-1}$. Higher velocities lead to an even larger broadening of the line so that they fall below the sensitivity %%@
limit of SPI.
A hexagonal dither pattern would have yielded a higher detection significance for the same observation time, but with %%@
the point source on axis, the equations which have to be solved for a SPI image reconstruction are underdetermined. %%@
So the SPI data would be of limited use.  

In the case of a SN of type Ib/II $^{44}$Ti is coming from the innermost ejected material. So it is expected that %%@
$^{44}$Ti was ejected at lower speed compared to the mean expansion velocity of known SNR of about $4600 \rm \, km\, %%@
s^{-1}$ \citep{2000ApJ...545L..53H}. 
For a Doppler-broadened line with FWHM of 2.4 keV, which corresponds to an expansion velocity of about $4600 \rm \, %%@
km\, s^{-1}$, the $^{44}$Ti 78.4 keV line will be detectable at $\sim 3.5\, \sigma$ level for the highest flux %%@
depicted in Fig.\,\ref{fig:44Ti-sens}. The instrumental energy resolution of the germanium detectors is about 1.8 keV %%@
in this energy range, thus it should be possible to determine/constrain the Doppler broadening and the shape of 
the line.

From IBIS/ISGRI interesting results can be expected \citep{2001egru.conf..509L}, because it provides a high angular %%@
resolution of 12 arcmin, which is high enough to spatially resolve GRO J0852-4642, but with the drawback of a %%@
degraded sensitivity  for sources which are not point like. Furthermore the energy resolution of ISGRI ($\Delta E/E$: %%@
9\% at 100 keV) is not sufficient to address a possible line broadening.
Already from the AO-1 observation of the Galatic center region with ISGRI upper limits on the $^{44}$Ti line flux at %%@
67.9 keV and 78.4 keV are presented by \cite{2004ESA...Renaud}. From IBIS/PICsIT we do not expect a significant %%@
result at 1157 keV, because its line sensitivity at this energy is more than an order of magnitude worse compared to %%@
that of SPI.

\section{Analysis}

As has been shown in Section \ref{chapter:SensEst} the search for an astrophysical $^{44}$Ti-line flux is most %%@
promising at the 78.4~keV line, especially for broadened lines. The analysis method\footnote{The data were prepared %%@
with the {\mbox {\it INTEGRAL}} off-line scientific analysis package, version OSA 3.0, released by ISDC.} uses %%@
imaging in a narrow energy band around this line from 77 to 80 keV. For an assessment of systematics and artefacts, %%@
the analysis was redone in an adjacent narrow energy band between 72 and 75 keV. In order to find strong continuum %%@
sources, which could produce fake line sources in the two narrow energy bands, a  search for continuum sources was %%@
performed in two broader energy bands, lacking of strong background lines, one below the interesting range from 30 to %%@
50 keV and one above from 70 to 80 keV.  In Fig.\,\ref{fig:spectrum} these four energy ranges are marked on top of a %%@
typical spectrum of SPI's camera.

\begin{table*}[bot]
\centering
\caption[]{Source fluxes extracted with SPIROS in imaging mode. The location of three sources were fixed by using an %%@
input catalog for SPIROS. The other sources were found freeely by SPIROS by searching for up to five new sources in %%@
the observed region. The four energy ranges used for the analysis were already presented in Fig.\,\ref{fig:spectrum}. %%@
Two of the corresponding significance maps of the Vela region are shown in Fig.\,\ref{fig:44Ti-maps}\\
}
\label{tab:flux-table}
\begin{tabular}{lcccc}
%Row: 1
\hline
 \rule[-1ex]{0pt}{3.5ex} Source & \multicolumn{4}{c}{Flux [$10^{-4}\, \rm \gamma \, cm^{-2}\, s^{-1}$]}  \\
%Row: 2
\hline
\rule[-1ex]{0pt}{3.5ex}  & "continuum" & "line" & "$^{44}$Ti - line" & "continuum"\\
%\cline{2-5}
\rule[-1ex]{0pt}{3.5ex} \raisebox{1.5ex}[-1.5ex]{E - Range}  & 30 -- 50 keV & 72 -- 75 keV & 77 -- 80 keV & 80 -- 90 %%@
keV\\
\hline
\rule[-1ex]{0pt}{3.5ex}& \multicolumn{4}{l}{sources fixed:} \\
%Row: 3
\rule[-1ex]{0pt}{3.5ex} Vela X-1 & 145.5  $\pm$0.7  &0.6  $\pm$ 0.2& 0.4 $\pm$ 0.2  & 1.4 $\pm$ 0.3 \\
%Row: 4
\rule[-1ex]{0pt}{3.5ex} GS0836-429 & 46.04 $\pm$ 0.7 & 1.3 $\pm$ 0.2 & 1.2 $\pm$ 0.2  & 1.5  $\pm$  0.3\\
%Row: 5
\rule[-1ex]{0pt}{3.5ex} GROJ0852-4642 & --  & -- & 0.0 $\pm$ 0.2 &  -- \\
\rule[-1ex]{0pt}{3.5ex} & \multicolumn{4}{l}{free search for new sources:} \\
%Row: 6
\rule[-1ex]{0pt}{3.5ex} (Vela Pulsar) & --  & -- &  0.4 $\pm$ 0.2  &  1.9 $\pm$ 0.3  \\
%Row: 7
\rule[-1ex]{0pt}{3.5ex} "Spurious" detections up to & 16 $\pm$ 0.7 & 0.6 $\pm$ 0.2 &  0.7 $\pm$ 0.2  & 1.1 $\pm$ 0.3 %%@
\\
\hline
\end{tabular}
\end{table*}

\begin{figure}
\centering
\includegraphics[width=0.92\linewidth]{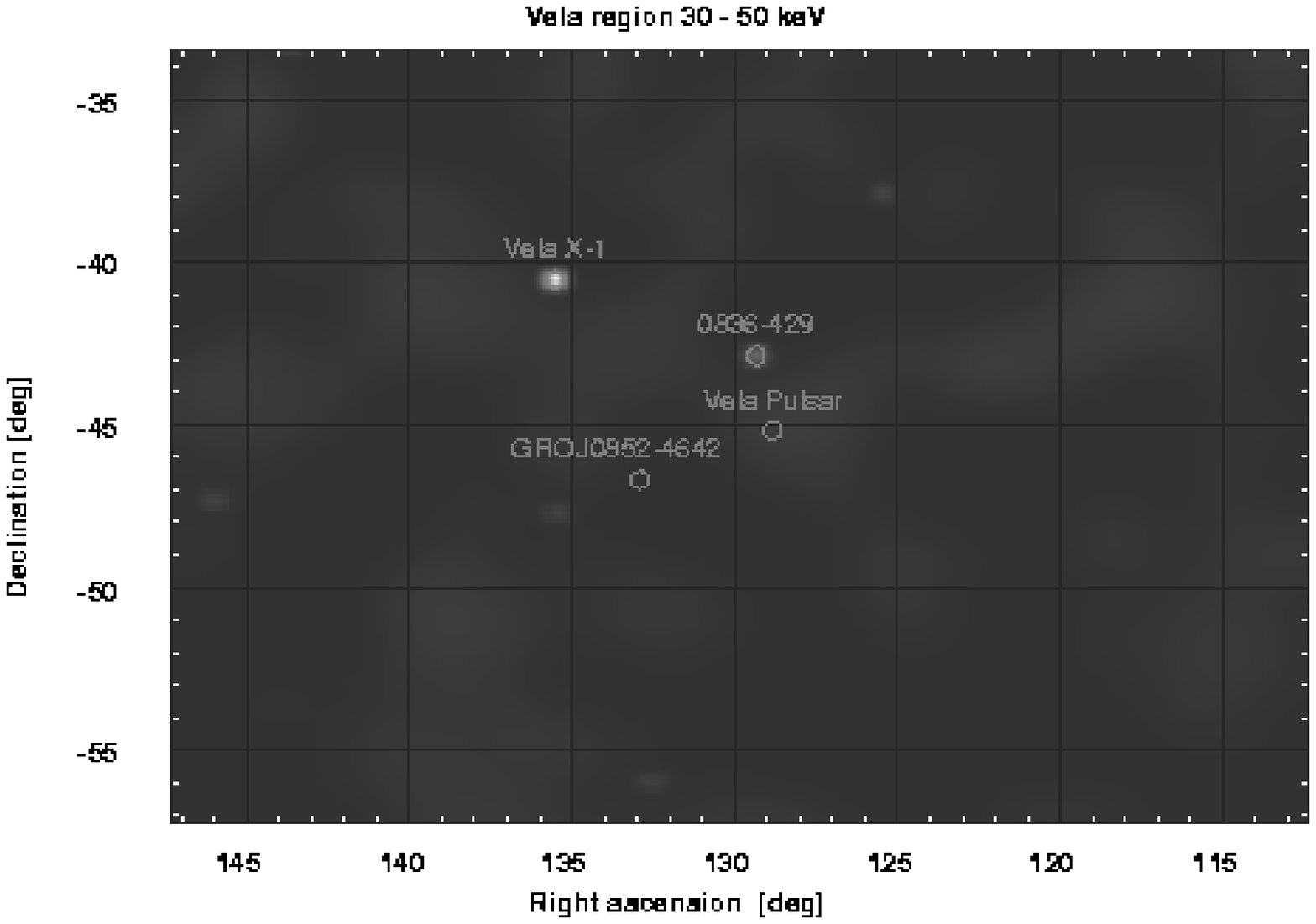}
\includegraphics[width=0.92\linewidth]{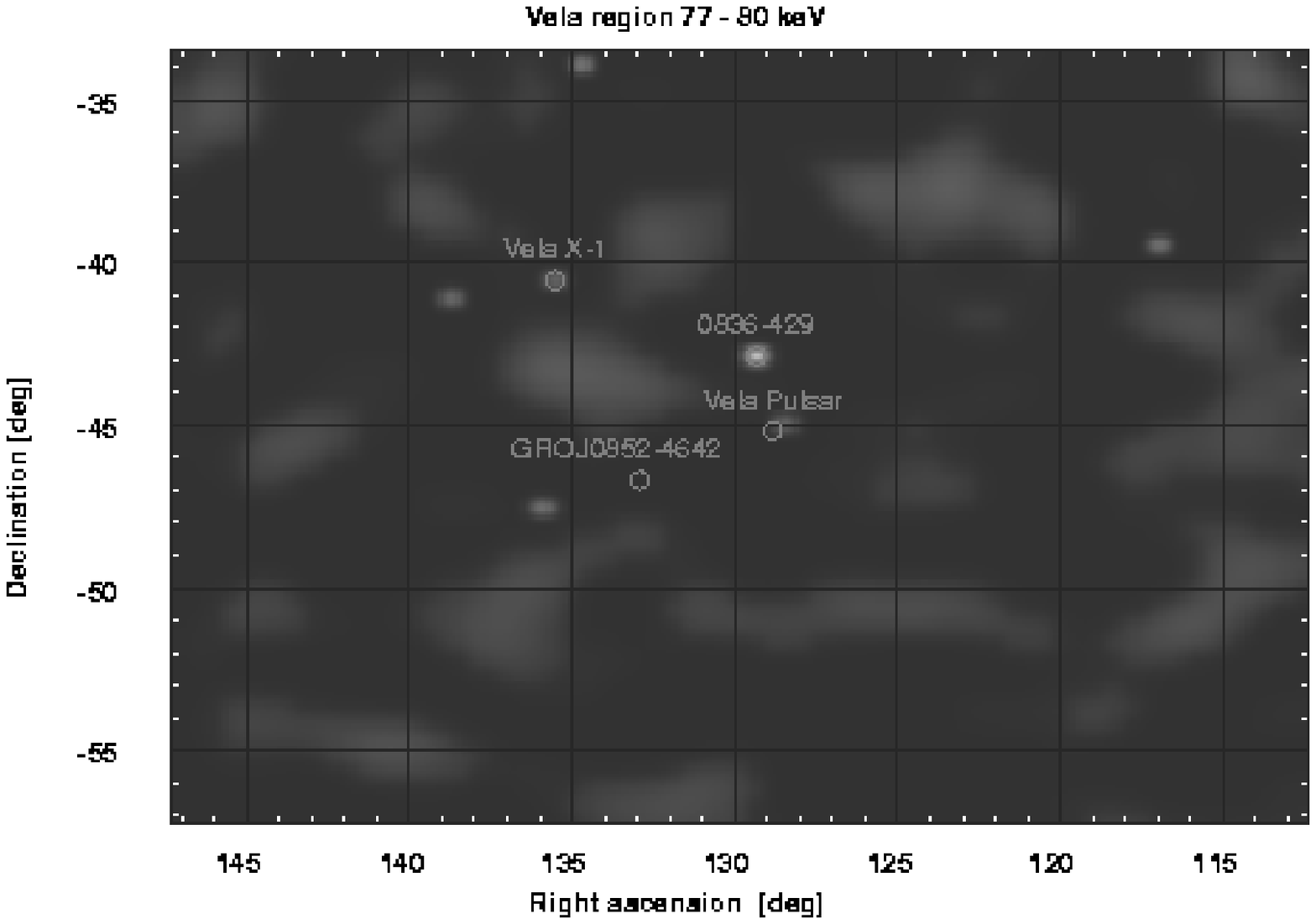}
\caption{SPIROS significance maps of the Vela region for the 30 and 50 keV energy band  (upper plot) and the 77 to 80 %%@
keV band (lower plot). The circles mark the positions of known sources. \label{fig:44Ti-maps}}
\end{figure} 

The imaging analysis was performed by using the SPIROS\footnote{SPIROS, version 6.0.} software %%@
\citep{2003A&A...411L.123S}. 
SPIROS allows the choice between different background models and methods. Commonly the saturated event rate of each %%@
germanium detector is used as a background tracer, which is a good measure of the high-energy particle flux impinging %%@
on SPI's camera. 
This rate can be used as an estimate of the background variation with time and  of the scaling factors between the %%@
detectors, which both will be fitted within SPIROS (SPIROS method 2). An alternative is the so called "mean %%@
count-rate method" (MCM: SPIROS method 5) which does not need a background model generated by another program like %%@
SPIBACK \citep{2003A&A...411L.117D}. SPIROS is calculating in this case the background internally by assuming that %%@
the measured rates of each detector are mainly due to background events. Their variations are accounted for changes %%@
of the background level. This method is not advisable in the case when strong or variable sources are in the field of %%@
view. 
The $\chi^2$ optimization statistics was used and pointings with extreme count residues above 3 $\sigma$ for each %%@
detector individually were automatically excluded by SPIROS. With this method it is possible to obtain  reduced %%@
$\chi^2$ values  $\leq 2$. 
The Vela region was analyzed with both background methods. Nevertheless a strong and variable source was in the field %%@
of view (Vela X-1), the results obtained with the MCM background method (summarized in Tab.\,\ref{tab:flux-table}) %%@
yielded  better $\chi^2$ values.

The results of the imaging analysis are summarized in Tab.\,\ref{tab:flux-table}. In all 4 energy bands the position %%@
of three sources, Vela X-1, the Burster GS0836-429 \citep{1992PASJ...44..641A} and the source of interest %%@
GROJ0852-4642 were fixed. With SPIROS a search for 5 new sources above a significance threshold of 3 $\sigma$ was %%@
performed and the fluxes of the fixed and newly found sources were extracted. Fig.\,\ref{fig:44Ti-maps} shows two of %%@
the obtained SPIROS detection significance maps. 
In both broad energy bands the two continuum sources  Vela X-1 and GS0836-429 were detected. In the 30 to 50 keV band %%@
with a high significance $ \sigma > 50$ and in the 80 to 90 keV band with $\sigma > 3 $.  In the two line-like energy %%@
bands only GS0836-429 was visible with $\sigma \geq 6$. The Vela Pulsar was found without catalogue input in the %%@
80-90\,keV energy band  at $\sim 6\, \sigma$, 0.33$^{\circ}$ off the catalogue position. No flux was detected from %%@
GROJ0852-4642 in any of the 4 energy bands, especially in the energy band where the $^{44}$Ti line flux is expected. %%@
In this case a statistical $2 \sigma$ upper limit of $4 \times 10^{-5}\, \rm \gamma \, cm^{-2}\, s^{-1}$ can be %%@
deduced.
In all four energy bands always 5 new sources, most of them obviously not real (except Vela pulsar), at random %%@
positions were found. The extracted flux levels of the strongest of these "spurious" sources is listed in the last %%@
row of Tab.\,\ref{tab:flux-table}.  A detailed discussion on the detection of spurious sources can be found in %%@
\cite{2004ESA...Dubath}

The occurrence of spurious sources can be used as a measure of the systematic uncertainties introduced by a not %%@
optimal handling of the background. For the $^{44}$Ti line energy band we detected spurious sources up to a flux of %%@
$7 \times 10^{-5}\, \rm \gamma \, cm^{-2}\, s^{-1}$. Adding  the statistical $2 \sigma$ upper limit, one gets a %%@
conservative upper limit for the $^{44}$Ti line flux at 78.4 keV for GROJ0852-4642 of $1.1 \times 10^{-4}\, \rm %%@
\gamma \, cm^{-2}\, s^{-1}$.

\begin{figure}[h]
\centering
\includegraphics[width=1.0\linewidth]{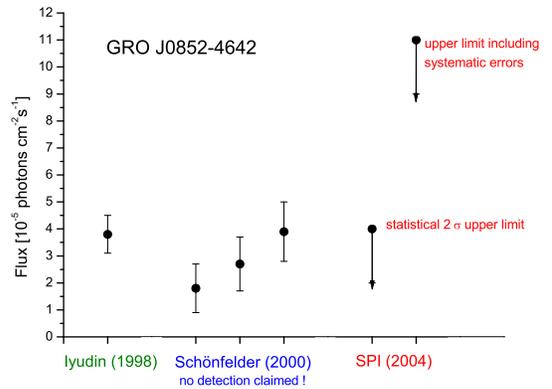}
\caption{SPI upper limit in comparision to literature values for the $^{44}$Ti line flux from GROJ0852-4642. %%@
\label{fig:Comparision_44Ti-flux}}
\end{figure}

\section{Conclusion}

In Fig.\,\ref{fig:Comparision_44Ti-flux} the upper limits derived from the current analysis of the SPI Vela %%@
observation are compared with flux values quoted for GROJ0852-4642 in the literature %%@
\citep{1998Natur.396..142I,2000AIPC..510...54S}. Without improving the systematic uncertainties the SPI result will %%@
not constrain the COMPTEL measurements.

\section*{Acknowledgments}
The SPI/{\mbox {\it INTEGRAL}} project is supported by the German "Ministerium f\"ur Bildung und Forschung" through 
DLR grant 50.OG.9503.0.

 \bibliographystyle{aa}
   \bibliography{vKie-ref}

\end{document}